\documentclass{appolb}

\usepackage{amssymb}
\usepackage{amsmath}
\usepackage{epsfig}
\usepackage{amsfonts}
\usepackage{graphicx}%

\begin{document}
\title{Lattice versus 2PI: 2d $O(N)$ model at nonzero $T$%
\thanks{Presented at the Workshop \textquotedblleft Excited QCD 2012\textquotedblright\ 
, 7-12 May 2012-Peniche }%
}
\author{Elina Seel$^{a}$, Dominik Smith$^{a,b}$, Stefano Lottini$^{a}$, \\
Francesco Giacosa$^{a}$
\address{$^{a}$Institut f\"ur Theoretische Physik, Johann Wolfgang Goethe Universit\"at,
Max-von-Laue-Str.\ 1, D--60438 Frankfurt am Main, Germany\\ 
$^{b}$ Fakult\"at f\"ur Physik, Universit\"at Bielefeld,\\
Universit\"atsstr.\ 25, D--33615 Bielefeld, Germany}
}
\maketitle
\begin{abstract}
The thermodynamics of the O(N) model in 1+1 dimensions is studied applying the CJT 
formalism and the auxiliary field method as well as fully nonperturbative finite 
temperature lattice simulations. The numerical results for the renormalized mass of 
the scalar particles, the pressure and the trace anomaly are presented and compared 
with the results from lattice simulation of the model. We find that when going to the 
two loop order we observe a good correspondence between the CJT formalism 
and the lattice study.
\end{abstract}
\PACS{11.10.Wx, 11.30.Rd, 12.39.Fe.}

\section{Introduction}

The two dimensional $O(N)$ nonlinear sigma model has many interesting features
in common with four-dimensional non-Abelian gauge theories, which makes it
useful to study as a toy model for QCD \cite{Novikov:1984ac, Rim:1984hk, 
Flyvbjerg:1989gk, Root:1974zr,Andersen:2004ae}. For instance, the coupling constant is
dimensionless, therefore the theory is renormalizable. Besides, this model is
asymptotically free and has a dynamically generated mass gap. Another
interesting property is the conformal invariance: In two dimensions the nonlinear
sigma model is classically scale invariant. However, on the quantum level a
scale is introduced due to renormalization of the quantum corrections.
Furthermore, for $N=3$ the model\ exhibits instanton solutions.
We start with the usual expression for the generating functional at nonzero
temperature%
\begin{equation}
\ Z=\mathcal{N}\int\mathcal{D}\Phi\delta\left(  \Phi^{2}-\frac{N}{g^{2}}\right)
\exp\left[  \overset{\ }{-\int_{0}^{\beta}}d\tau\int_{-\infty}^{\ \infty
}dx\mathcal{L}_{0}\right]  \text{ ,}%
\end{equation}
where $g$ is the coupling constant and $\mathcal{L}_{0}$ is a free Lagrangian
\begin{equation}
\mathcal{L}_{0}=\dfrac{1}{2}\partial_{\mu}\Phi^{t}\partial_{\mu}%
\Phi,\ \ \ \ \Phi^{2}=\Phi^{t}\Phi,\Phi^{t}=\left(  \sigma,\pi_{1},\ldots
,\pi_{N-1}\right) \text{ .}
\end{equation}
The fields are restrained by the
condition $\Phi^{2}=N/g^{2}$ which is incorporated by the delta function. The
nonlinear constraint enforces the thermodynamics of the model on an $N-1$
dimensional hypersphere and induces the interactions between the fields. Using
the mathematically well-defined (i.e., convergent) form of the usual
representation of the functional $\delta$-function%
\[
\delta\left(  \Phi^{2}-\dfrac{N}{g^{2}}\right)  =\lim_{\varepsilon
\rightarrow\text{ }0^{+}}N\int\mathcal{D}\alpha e^{\left\{  -\ \overset
{\ }{\int_{0}^{\beta}}d\tau\int_{-\infty}^{\ \infty}dx\left[  \frac{i\alpha
}{2}\left(  \Phi^{2}-\frac{N}{g^{2}}\right)  +\frac{\varepsilon\alpha^{2}}%
{2}\right]  \right\}  }\text{ }%
\]
the generating functional and the corresponding Lagrangian of the $O(N)$
nonlinear model can be rewritten as follows \cite{non-lin} 
\begin{align*}
\ Z &  =\lim_{\varepsilon\rightarrow\text{ }0^{+}}\mathcal{N}\int\mathcal{D}%
\alpha\mathcal{D}\Phi\exp\left[ \overset{\ }{-\int_{0}^{\beta}}d\tau
\int_{\infty}^{\ \infty}dx\mathcal{L}\right]  \text{ },\\
\mathcal{L} &  =\dfrac{1}{2}\partial_{\mu}\Phi^{t}\partial_{\mu}\Phi
+U(\Phi,\alpha)\text{ , \ \ \ }U(\Phi,\alpha)=\dfrac{i}{2}\alpha(\Phi
^{2}-\dfrac{N}{g^{2}})+\dfrac{\varepsilon}{2}\alpha^{2}\text{ ,}%
\end{align*}
where $\alpha$ is an auxiliary field serving as a Lagrange multiplier.

\section{Analytic calculations}
In order to study the thermodynamic behavior we apply the CJT formalism. Within
this formalism we obtain the following expressions for the renormalized
effective potential and for the renormalized gap equation to one-loop order

\begin{align*}
V_{eff}^{ren}  & =N\int_{0}^{\infty}\dfrac{dk}{\pi}\dfrac{k^{2}}{\omega_{k}%
}\dfrac{1}{\exp\left\{  \omega_{k}/T\right\}  -1}+NM^{2}\left[  \left(
\dfrac{1}{2g_{ren}^{2}}-\frac{1}{4\pi}\left(  1+\ln\frac{\mu^{2}}{M^{2}%
}\right)  \right)  \right]  \text{ },\\
\frac{1}{g_{ren}^{2}}  & =\int_{0}^{\infty}\dfrac{dk}{\pi}\dfrac{1}{\omega
_{k}}\dfrac{1}{\exp\left\{  \omega_{k}/T\right\}  -1}+\frac{1}{4\pi}\ln
\frac{\mu^{2}}{M^{2}}\text{ ,}%
\end{align*}
where $\mu$ is the renormalization parameter, $m$ is the vacuum mass, \ $M$ is
the temperature dependent mass and $g_{ren}^{2}$ is the renormalized coupling constant.
The details of the computation can be found in \cite{O3}.
Next, we can calculate the thermodynamic pressure which is, up to a minus sign,
identical to the minimum of the effective potential
\[
P=-V_{eff}^{min}\text{ .}%
\]
As was shown in \cite{Andersen:2004ae}, to two-loop order one must apply numerical
methods in order to regularize the effective potential at finite temperature.
The final result is rather lengthy and is given in \cite{Andersen:2004ae} as well 
as in \cite{O3}.

\section{Lattice simulation}

\begin{figure}
[ptb]
\begin{center}
\includegraphics[
width=4in
]%
{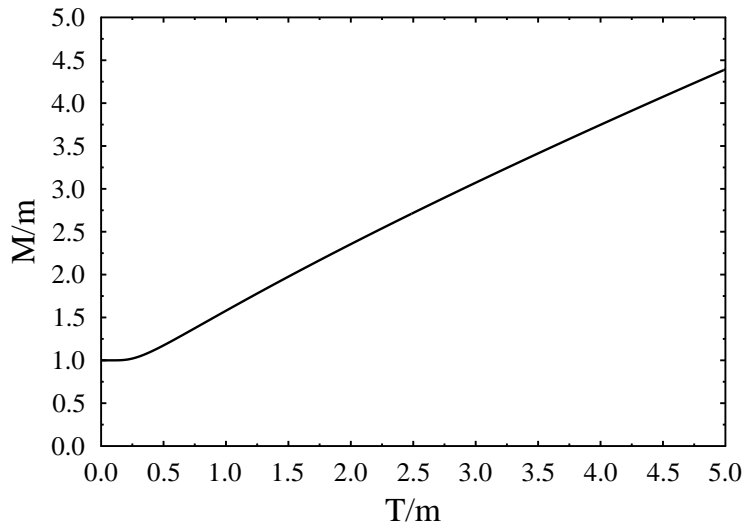}%
\caption{The analytic result for the mass of the scalar field as a function
of $T/m.$}%
\label{m}%
\end{center}
\end{figure}
In this section we summarize the thermodynamic approach applied by finite
temperature lattice simulations. The Euclidean, discretised action takes the
form of a Heisenberg model,
\begin{equation}
S=\beta\sum_{\langle i,j\rangle}\left(  1-\vec{s}_{i}\cdot\vec{s}_{j}\right)
\;,\label{eq:s_latt}%
\end{equation}
where the sum runs over all bonds of a 2-dimensional lattice, $\vec{s}_{i}$
are 3-dimensional unit vectors in internal space and $\beta=N/g^{2}.$ The
corresponding partition function reads%
\begin{equation}
\mathcal{Z}\propto\Bigg(\prod_{k}\int_{S^{2}}\vec{s}_{k}\Bigg)\,\prod_{\langle
i,j\rangle}e^{\beta\vec{s}_{i}\cdot\vec{s}_{j}}\;.
\end{equation}
The system at finite temperature $T$ is realized by making the time-like extent
of the lattice finite and consisting of $N_{t}$ sites, with periodic boundary
conditions in that direction; denoting with $a$ the lattice spacing, we have
$aN_{t}=1/T$. 

In order to evaluate the pressure we use
the \textquotedblleft integral method\textquotedblright\ \cite{Boyd:1996bx}
and  additive renormalization
\begin{equation}
\frac{p(T)}{T^{2}}=N_{t}^{2}\int_{0}^{\beta}\left(  \left\langle \ell
_{x}+\ell_{t}\right\rangle _{\beta^{\prime},N_{t}}-2\left\langle
\ell\right\rangle _{\beta^{\prime},\infty}\right)  d\beta^{\prime
}\;,\label{eq:presint}
\end{equation}
where $\ell_{e}=\vec{s}_{i}\cdot\vec{s}_{i+\hat{e}}$.  This requires the
knowledge of the \textquotedblleft beta function\textquotedblright\
$\partial\beta/\partial\ln T=-a\partial\beta/\partial a$, which is extracted 
nonperturbatively, measuring the running of the coupling
with the scale and using a suitable parametrization of the data. Note that the
system at $T=0$ is numerically approximated with a large enough square
system, that is $N_{x}=N_{t}$.

The running of the scale $a(\beta)$ is needed to convert the temperature in
physical units. This is realized by computing the spin-spin correlation
function in a zero-temperature system over a wide range of couplings 
$C(an)=\langle\vec{s}_{i}\cdot\vec{s}_{i+n\hat{e}}\rangle,$ from which
we obtain the zero-temperature mass $m$. Distances on the lattice
are in units of $a$, therefore the masses we measure are the adimensional
quantities $am$; once we fix the corresponding physical value we have the
function $a(\beta)$ in units of $1/m$.

\section{Results and Discussion}

\begin{figure}
[ptb]
\begin{center}
\includegraphics[
width=4in
]%
{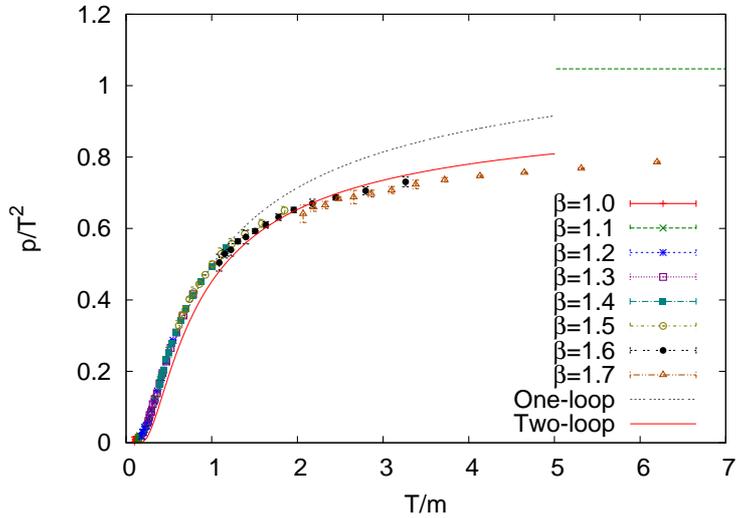}%
\caption{The pressure as a function of $T/m$ in one-loop (blue) and in
two-loop (red) approximation compared to lattice simulations for different
values of $\beta=N/g^{2}$.}%
\label{p}%
\end{center}
\end{figure}

In this subsection we compare the analytic results with lattice simulations.
We choose $N=3$, corresponding to a system of three scalar fields. The vacuum
mass of the scalar field $m$ is the only dimensionful parameter of the model.
Therefore we plot all thermodynamic observables in units of $m.$

Figure \ref{m} shows the analytic result for the temperature dependent
renormalized mass of the scalar fields. An important observation is that it
has a similar behavior as the gluon mass in the deconfined phase, 
e.g. ref. \cite{fran} and refs. therein. The model
exhibits dimensional transmutation just like QCD, meaning that at zero
temperature there is a nonvanishing mass gap, which\ is generated due to
renormalization of quantum corrections. Besides, at high $T$ the temperature
dependence of the mass can be approximately parametrized by $T/\log T.$ This
is exactly the same behavior that one observes for the gluon mass in the
deconfined phase.

Figure \ref{p} shows the thermodynamic pressure. The uncertainties 
are very small and completely overcome by systematics. The high temperature 
limit is better described in the two-loop
calculations, whereas at low temperatures there is a better correspondence
between the lattice simulations and the one-loop calculations. We find that at
very high temperatures one degree of freedom becomes effectively removed, and
the pressure approaches the limit of a non-interacting gas of $N-1$ bosons,
$P/N=(N-1)\pi T^{2}/6N,$ which is indicated by the straight dashed line 
in the upper right corner. This is an immediate consequence of the asymptotic 
freedom and of the nonlinear
constraint. One can understand this behavior by remembering that we start
with a free Lagrangian for $N$ scalar fields. Introducing the nonlinear
condition, $\Phi^{2}=N/g^{2}$, the thermodynamics is constraint on a two
dimensional sphere which effectively removes one degree of freedom.
Analytically this removal of one degree of freedom can only be described in
the 2-loop approximation.

\end{document}